\documentclass[fleqn,usenatbib]{mnras}

\usepackage[T1]{fontenc}
\usepackage{ae,aecompl}

\usepackage{graphicx}	
\usepackage{amsmath}	
\usepackage{amssymb}	

\renewcommand {\deg}   {\mbox{$^\circ$}}

\newcommand   {\kms}   {\mbox{km\,s$^{-1}$}}
\renewcommand {\ga}    {\mbox{\rlap{\hbox{\lower5pt\hbox{$\sim$}}}\hbox{$>$}}}
\renewcommand {\la}    {\mbox{\rlap{\hbox{\lower5pt\hbox{$\sim$}}}\hbox{$<$}}}

\newcommand{\ee}[1]	{\times 10$^{#1}$} 

\title[Origin of NRFs]{G0.173-0.42: an X-ray and radio
magnetized filament near  the galactic center}

\author[F. Yusef-Zadeh, M. Wardle, C. Heinke, R. Arendt, M. Royster,  I. Heywood, W. Cotton,  F. Camilo \& J. Michail]
{F. Yusef-Zadeh$^1$\thanks{E-mail: zadeh@northwestern.edu}, M. Wardle$^{2}$,
C. Heinke$^3$, I. Heywood$^{4,5,6}$, R. Arendt$^{7}$,
\newauthor M. Royster$^1$, W. Cotton$^8$,  F. Camilo$^6$ \& J. Michail$^1$\\
$^{1}$CIERA, Department of Physics and Astronomy Northwestern University, Evanston, IL 60208\\
$^{2}$Dept of Physics and Astronomy,  Research Centre for Astronomy, Astrophysics\\
and Astrophotonics, Macquarie University, Sydney NSW 2109, Australia\\
$^{3}$Dept of Physics, University of Alberta, CCIS-4-183, Edmonton, AB T6G 2E1, Canada\\ 
$^{4}$Astrophysics, Department of Physics, University of Oxford, Keble Road, Oxford, OX1 3RH, UK\\
$^{5}$Department of Physics and Electronics, Rhodes University, PO Box 94, Makhanda, 6140, South Africa\\
$^{6}$South African Radio Astronomical Observatory, 2 Fir Street, Black River Park, Observatory, 
Cape Town, 7925, South Africa\\
$^{7}$UMBC/GSFC/CRESST 2, Code 665, NASA/GSFC, 8800 Greenbelt Rd, Greenbelt MD 20771\\
$^{8}$National Radio Astronomy Observatory, Charlottesville, VA, USA}

\date{Accepted XXX. Received YYY; in original form ZZZ}

\pubyear{2019}

\begin{document}
\label{firstpage}
\pagerange{\pageref{firstpage}--\pageref{lastpage}}
\maketitle
\def\msol{\hbox{$\hbox{M}_\odot$}}
\def\lsol{\hbox{$\hbox{L}_\odot$}}
\def\kms{km s$^{-1}$}
\def\Blos{B$_{\rm los}$}
\def\etal   {{\it et al.}}                     
\def\psec           {$.\negthinspace^{s}$}
\def\pasec          {$.\negthinspace^{\prime\prime}$}
\def\pdeg           {$.\kern-.25em ^{^\circ}$}
\def\degree{\ifmmode{^\circ} \else{$^\circ$}\fi}
\def\ut #1 #2 { \, \textrm{#1}^{#2}} 
\def\u #1 { \, \textrm{#1}}          
\def\nH {n_\mathrm{H}}
\def\ddeg   {\hbox{$.\!\!^\circ$}}              
\def\deg    {$^{\circ}$}                        
\def\le     {$\leq$}                            
\def\sec    {$^{\rm s}$}                        
\def\msol   {\hbox{$M_\odot$}}                  
\def\i      {\hbox{\it I}}                      
\def\v      {\hbox{\it V}}                      
\def\dasec  {\hbox{$.\!\!^{\prime\prime}$}}     
\def\asec   {$^{\prime\prime}$}                 
\def\dasec  {\hbox{$.\!\!^{\prime\prime}$}}     
\def\dsec   {\hbox{$.\!\!^{\rm s}$}}            
\def\min    {$^{\rm m}$}                        
\def\hour   {$^{\rm h}$}                        
\def\amin   {$^{\prime}$}                       
\def\lsol{\, \hbox{$\hbox{L}_\odot$}}
\def\sec    {$^{\rm s}$}                        
\def\etal   {{\it et al.}}                     
\def\la{\lower.4ex\hbox{$\;\buildrel <\over{\scriptstyle\sim}\;$}}
\def\ga{\lower.4ex\hbox{$\;\buildrel >\over{\scriptstyle\sim}\;$}}
\def\ee #1 {\times 10^{#1}}          

\begin{abstract} 
The recent detection of an X-ray filament associated with the radio filament G$0.173-0.42$ adds to four other 
nonthermal radio filaments with X-ray counterparts, amongst the  more than 100 elongated radio structures that have been 
identified as synchrotron-emitting radio filaments in the inner couple of degrees of the Galactic center. 
The synchrotron mechanism has also been proposed to explain the emission from 
X-ray filaments.  However, the origin of radio filaments and the acceleration sites of energetic particles to produce 
synchrotron emission in radio and X-rays remain mysterious. Using  MeerKAT, VLA, {\it Chandra}, {\it WISE} and {\it 
Spitzer}, we present structural details of G0.173-0.42 which consists of multiple radio filaments, one of which has an 
X-ray counterpart. A faint oblique radio filament crosses the radio and X-ray filaments. Based on the morphology, 
brightening of radio and X-ray intensities, and radio spectral index variation, we argue that a physical interaction is 
taking place between two magnetized filaments. We consider that the reconnection of the magnetic field lines at the 
interaction site leads to the acceleration of particles to GeV energies. We also argue against the synchrotron 
mechanism for the X-ray emission due to the short $\sim$30 year lifetime of TeV relativistic particles. Instead, we 
propose that the inverse Compton scattering mechanism is more likely to explain the X-ray emission by upscattering of 
seed photons emitted from a $10^6$ \lsol\, star located at the northern tip of the X-ray filament. 
\end{abstract}



\section{Introduction} 


More than 30 years have elapsed since the nonthermal radio filaments associated with the Galactic center 
radio Arc near $l\sim0.2^\circ$ were first reported \citep{zadeh84}. These observations showed linear, 
magnetized features running perpendicular to the Galactic plane and since then more than 100 nonthermal 
radio filaments (NRFs) with similar characteristics  have been discovered 
\citep{liszt85,zadeh86,morris89,gray91,haynes92,lang99,larosa04,zadeh04,law08,heywood19}. The intrinsic 
polarization observed from these filaments shows that their magnetic fields are directed along the filaments \citep{zadeh97,pare19}. 
The mechanisms responsible for accelerating particles to 
relativistic energies and for creating the  elongated filamentary geometry are still mysterious.
While several authors have examined how 
the filaments might arise from their interaction with molecular and ionized clouds or with mass-losing 
stars there is no consensus on any model \citep{rosner96,shore99,bicknell01,zadeh19}.

{\it Chandra}, {\it XMM} and {\it NuSTAR}
 have detected X-ray emission from a handful of nonthermal radio filaments. There are four prominent radio and X-ray 
filaments that have been studied in detail, G359.89--0.08 (Sgr A-E), G359.54+0.18 (ripple), G359.90-0.06 (Sgr A-F), G0.13-0.11, 
\cite{sakano03,lu03,lu08,zadeh05,zhang14,zhang20}. 
Unlike the long  and distinct radio  filaments, 
there are also several short, linear X-ray features that are identified in the inner 6$'$ 
of Sgr A* \citep{muno08,johnson09,lu08}.  
No detailed, high-resolution studies of the radio counterparts to these X-ray filaments have been published.
In all previous studies, 
the synchrotron scenario  for the X-ray emission has  been proposed with the exception of G359.90-0.06 (Sgr A-F)  
which could be explained either by synchrotron or the inverse Compton scattering (ICS) \citep{zadeh05}.     
 There are limited 
studies  investigating  the spectrum of the emission between radio and X-rays to determine the emission
mechanism in radio and X-rays 
\citep{zadeh05}.  Furthermore, the origin of X-ray filaments in many studies 
are speculated to be traces of  pulsar wind nebulae associated with pulsars 
\citep{lu03,lu08,zhang20}. 

Here we focus on  a newly discovered source, an X-ray counterpart \citet{zhang20a}
to the  radio filament G0.173-0.42, which is also called G0.17-0.42 or S5 in \citep{zadeh04}.  
This prominent radio 
filament  lies towards  eastern boundary of a diffuse, large-scale linearly polarized plume-like structure that runs 
toward negative latitudes of the Galactic plane \citep{zadeh86,zadeh90,seiradakis85,tsuboi95,zadeh04}. 
G0.173-0.42 consists of two parallel filaments with an extent of $\sim11'$ oriented perpendicular to the 
Galactic plane \citep{zadeh04}. We present X-ray observations of this filament indicating 
X-ray emission with an extent of $\sim2'$ along the 
radio filament. 

A faint oblique radio filament crosses the radio and X-ray filament.  We suggest that the acceleration of 
relativistic particles to GeV energies occurs due to reconnection of the magnetic fields at the location where the 
oblique filament crosses G0.173-0.42. We argue that the flow from the acceleration site encounters a luminous 
mass-losing star, thus the flow wraps around the mass-losing envelope of the star before it continues to the south.  
In this picture, the ICS is generating X-ray emission.

\section{Observations and Data Reduction} 

\subsection{VLA radio observations}

Radio continuum observations at 20 and 6cm (project AY15) were carried out using the VLA 
in its hybrid BnC- and CnD-array 
configurations on October 6, 1986 and February 7, 1987, respectively. These scaled-array 
observations were made to  determine the spectral index of G0.173-0.42. These 
observation were carried out in full polarization and narrow 2$\times50$ MHz band. We used 3C286 to 
calibrate the flux density scale, and 1748-253 to calibrate the complex gains. We used three overlapping 
6cm fields; we are are focusing on one 6cm field where X-ray emission is detected.  The pointing centers 
at nominal frequencies at  4.80 and 1.47 GHz are  
$\alpha, \delta(J2000)=17^h 47^m 20^s.723, -28^\circ 59' 23''.362$
and 
$\alpha, \delta(J2000)=17^h 47^m 40^s.768, -29^\circ 01' 00''.806$, respectively.

\subsection{MeerKAT radio observations}
The Galactic center region was observed by MeerKAT as part of its commissioning phase. The pointing used 
here was observed for 10.8 hours on 15 June 2018, with the array pointing at 
$\alpha, \delta(J2000)=17^h 47^m 38^s.34, -29^\circ 06' 18''.95$, 
for an on-source time of 6.84 hours. A full overview of the project will be
provided by Heywood et al. (in prep.), however a brief description of the major data processing steps
are as follows. Averaging was applied to the data to reduce the native 4,096 channels by a factor of 4.
Basic flagging commands were applied using the {\tt flagdata} task in {\tt CASA} 
including bandpass edges and regions of persistent radio frequency interference. Delay and bandpass
corrections were derived from observations of the primary calibrator source PKS~B1934-638, which was
also used to set the absolute flux scale. Time-dependent gains were derived from observations of a
bright (8~Jy at 1.28~GHz) calibrator source 1827$-$360, which was observed for 1 minute for every 10
minute target scan. Gain corrections were derived iteratively with rounds of residual flagging in
between. Following the application of these corrections, the target data were flagged using the {\tt
tricolour} package\footnote{\url{https://github.com/ska-sa/tricolour}}, and then imaged using {\tt
wsclean} \citep{offringa14} with multiscale cleaning \citep{offringa17} and iterative threshold-based
masking. Phase-only self-calibration solutions were derived for every 128 seconds of data using the {\tt
gaincal} task in {\tt CASA}, and the imaging process was repeated. A \citet{briggs95} robustness
parameter of --1.5 was used to provide high angular resolution. The primary beam attenuation was
corrected for by dividing the image by an azimuthally-averaged Stokes I beam model evaluated at 1.28~GHz
using the {\tt eidos} software \citep{asad19}.

The spectral index $\alpha$, is defined as I$_\nu\propto\nu^{\alpha}$ where I$_\nu$ is the intensity.  Fifteen 
narrow channels  within the broad 20 cm band were used to  accurately determine 
the spectral index distribution. The procedure to make a spectral index image is as follows.  
Imaging  the data in 15 sub-bands follows  the application of a Gaussian taper   to the visibilities of each sub-band 
to attempt to force common angular resolution. 
An inner cut to the {\it{u,v}} plane 
is also applied, to prevent the lowest frequencies seeing large angular scales that are not visible at higher frequencies.
 Each sub-image is primary beam  corrected, and convolved with a circular Gaussian with a FWHM$\sim8''$.
The primary beam corrected, common resolution images are stacked into a cube, which is then masked below a threshold $\sim1$ mJy.
A linear fit for the gradient of the spectrum is made in log-frequency/log-flux space for every sight line through the masked cube 
for which more than 50\% of the frequency planes contain total intensity measurements.

\subsection{{\it Chandra} X-ray observations}

We use data from three {\it Chandra X-ray Observatory} observations, ObsIDs 7157 (14.9 ks), 19448 (44.5 ks), and 
20111 (14.9 ks). Each observation used the Advanced CCD Imaging Spectrometer (ACIS) in the ACIS-I configuration. 
The data were processed using the {\it Chandra} X-ray Center’s {\tt CIAO} software package. We created images using 
exposure maps that divide out the instrument effective area in a variety of energy ranges, and eventually adopted 
1.5-8 keV to create images (see Figs. 1b,d and 4a). 
We show images smoothed with a Gaussian with  FWHM = $4''\times4''$.   
The {\tt CIAO} software {\tt vtpdetect} detects the filament with a false alarm 
probability of $2\times10^{-33}$.  The filament is oriented at 2 degrees clockwise of north-south in Galactic 
coordinates.  The filament seems to extend for at least 100” ($b = -00^\circ\, 26'\, 00''$ to $ b = -00^\circ\, 24'\, 15''$). 
It may extend farther to Galactic north and Galactic south. 

Inspection of the merged image suggests that the filament shows three parts with different fluxes, which we 
designate North, South, and Center.  The Center portion has the highest flux per unit area. A possible point source 
may be present at 
$l =  00^\circ\, 10'\,  25''.61\,   b = -00^\circ\,  25'\,  15''.99$
We extract spectra from the filament around this spot, designating it the 
Center region, and from the filament on either side of it. We use extraction regions 7” in width, and with lengths 
$58.2'', 5.7''$, and $41.7''$, for the North, Center, and South portions respectively.  Background regions are extracted 
from regions of similar width and length offset to either side of the filament by $\sim10''$. The total number of 
counts in each region is 279, 42, and 138 counts respectively. After background subtraction, we infer 260$\pm$23 
counts altogether can be attributed to the filament.


We fit the spectrum of the filament as a whole, and separately as three individual parts, using {\tt XSPEC} version 
12.10.1f.  We do not combine spectra, but fit them together in {\tt XSPEC}, tying parameters when appropriate (thus, all 
parameters are tied between epochs when fitting the spectrum of the whole filament).  We fit power-law and thermal 
bremsstrahlung spectra, with interstellar absorption represented by the {\tt tbabs} model assuming the {\tt wilm} 
interstellar abundances of \citet{wilms01}. Due to the small number of counts, we bin the data to 1 count/bin and 
fit using the C-statistic.

For the full spectrum fit with a power-law, we find best-fit values of $N_H=12^{+10}_{-6}\times10^{22}$ cm$^{-2}$, 
photon index 
 $\Gamma=2.5^{+1.7}_{-1.2}$, and 2-10 keV fluxes of (unabsorbed) $2.0^{+2.5}_{-0.5}\times10^{-13}$ ergs s$^{-1}$ 
 cm$^{-2}$, or (absorbed) $1.0^{+0.2}_{-0.3}\times10^{-13}$ ergs s$^{-1}$ cm$^{-2}$.  Extrapolating this fit to 
 0.5-10 keV, we find an unabsorbed flux of 
$5.5\times10^{-13}$ ergs s$^{-1}$ cm$^{-2}$ (the high 
 absorption and uncertainty in photon index make the intrinsic flux below 2 keV poorly constrained). Simulating 
 1000 fake datasets using the {\tt goodness} command, we find that 20\% give smaller test statistic values (using 
 the Cramer-von Mises statistic), so the power-law model is acceptable.


We also test a thermal bremsstrahlung model, for which we find $N_H=12^{+10}_{-6}$ cm$^{-2}$, $kT=5^{+81}_{-3}$ 
keV, and unabsorbed (absorbed) 2-10 keV flux of $1.7^{+0.8}_{-0.4}\times10^{-13}$ 
($1.0^{+0.1}_{-0.2}\times10^{-13}$) ergs s$^{-1}$ cm$^{-2}$. As the goodness for the bremsstrahlung fit is 24\%, we 
cannot discriminate between these two models.

Fitting the three spatial parts of the X-ray filament with a power-law, we try either tying $N_H$ and $\Gamma$ 
between the parts, or just tying $N_H$. For the model with tied $\Gamma$, we find 2-10 keV unabsorbed fluxes of 
$1.4^{+1.6}_{-0.4}\times10^{-13}$, $0.2^{+0.3}_{-0.1}\times10^{-13}$, and $0.4^{+0.6}_{-0.2}\times10^{-13}$ ergs 
s$^{-1}$ cm$^{-2}$, for the North, Center and South portions respectively. If the power-law indices are freed, we 
find values of $\Gamma=2.9^{+1.8}_{-1.4}$, $2.5^{+2.2}_{-2.0}$, and $1.8^{+1.9}_{-1.8}$ for the North, Center, and 
South portions respectively. We thus cannot distinguish whether there is any variation in power-law index among the 
three components.

\subsection{2MASS, {\it Spitzer} and {\it WISE}}

In addition to radio and X-ray data, we have also examined 2MASS,  {\it Spitzer} (IRAC and MIPS) and {\it WISE} 
images and point source  catalogs of the Galactic center \citep{skrutskie06,stolovy06,ramirez08,zadeh09}. 
We employed these data to investigate the SEDs of sources that appear to be 
associated with filaments.

\section{Results} 

\subsection{Morphology}

Figure 1 shows a close-up view of G0.173-0.42 in four different panels at 20 cm, 2-8 keV, 3.6 $\mu$m and 
composite 3-color image of radio, X-ray and infrared. Figure 1a shows the eastern filament 
with a peak flux density of $\sim0.75$ mJy beam$^{-1}$
at the position A,  $l = 10'\, 24.3'', b = -24'\, 53.70''$,  
where the oblique filament  crosses the parallel filaments 
($''\rm A''$ is defined as the intersection of the eastern filament with the oblique filament).  
When compared to typical surface brightness of G0.173-0.42, 
the intensity of  the eastern filament increases  by a factor of 2-3 near position A. 
Enhanced radio emission extends to both 
north and south of position A along the eastern filament. 
Figure 1b shows  the X-ray emission  near position A
traces enhanced radio emission along the eastern filament. 

At its north end, the X-ray filament deviates to the northeast and breaks up into two filaments. This is coincident 
with a luminous infrared star located at $l=10' 18''.81$, $b=23' 44''.61$ (circled in the 3.6 $\mu$m image in Figure 
1c). The star may be the brightest member of a cluster. The candidate cluster is defined by the 6 stars with $[3.6] < 
8.5$ mag that are located within $20''$ of the bright star. There are no other groupings as bright and dense as this 
within $200''$, implying a less than 
1\% chance that the candidate cluster is simply a random arrangement of stars on 
the line of sight.
The X-ray filament deviates  to the 
northeast and  breaks up into diffuse circular-shaped structure  where a stellar cluster at 3.6$\mu$m is found. 
Figure 1d shows a composite image revealing that the 
northern tip  of X-ray filament has no radio counterpart
but coincides with the 3.6$\mu$m  emission from the  stellar cluster.

The comparison of radio and X-ray images shows that the X-ray counterpart to G0.173-0.42 
lies to the east of the radio filament and the western radio filament has no X-ray counterpart (see Fig. 1c). 
We have not astrometrically corrected the radio and X-ray images to each other. 
The 90\% uncertainty absolute astrometry errors for Chandra data are 0.8”\footnote{ 
https://cxc.cfa.harvard.edu/cal/ASPECT/celmon/}, while radio interferometric astrometric errors 
are comparably negligible.
For the VLA,  the astrometric accuracy  is usually down 
 to a few percent of the beamsize.  
MeerKAT data give  systematic errors up to an arcsecond. 
A shift of $1''$ (arcsecond)  in the X-ray or radio  positions would not affect the relative alignment of the filaments 
which are several arcseconds across. 

The X-ray filament has a length  of $\sim2'-2.5'$, far less than the radio filament that extends 
for $\sim11'$ \citep{zadeh04}. A second  filament also runs parallel to G0.173-0.42 and the two filaments 
extend for a total of $\sim20'$ (or 48pc at the 8 kpc Galactic  center distance), as shown in Figure 2. 
We note that the strongest emission  in Figure 2 arises from 
G0.173-0.42 where X-ray emission is detected. 
Lastly, 
the northern tip  of the X-ray filament near $b\sim-23'\, 45''$  deviates from radio filaments that themselves bend at 
Galactic latitude $b\sim-23'$ by about 6 $arcdeg$ to the northwest.   
The  high resolution MeerKAT 20 cm image of G0.173-0.42 reveals the east and west filaments have 
 the appearance of winding about each other and converging to the south near $b\sim-27'$. 
We note a 
third filament G0.167-0.405 crossing G0.173-0.42 at position A at an oblique angle of $-32^\circ$. The intensity of radio emission is 
stronger by a factor 2-3 to the south of A when compared to the north of A. 
As described below, radio  spectrum  becomes increasingly flatter to the south of position A.

Figures 1b and 1d  show the X-ray filament encounters
the stellar cluster. It appears that the northern extension of the  X-ray  filament 
from position A  splits 
and wraps around the brightest member of the cluster 
and continues as two faint parallel linear features before the emission is terminated. 
We consider below that there is a physical  interaction 
of the X-ray filament with the envelope this (presumed) mass-losing luminous star. 
Figure 3 shows the SED of the bright star derived from  2MASS, IRAC, MIPS  {\it WISE} data. 
This star IRAC flux densities near or above the expected saturation limits. 
A fit to the 2MASS and {\it WISE} SED is also shown.  
Using the {\it WISE} data (black squares)  instead of the saturated {\it Spitzer} IRAC data (open circles), 
the SED suggests about 2 mag extinction at K band with the corrected SED (red squares) characterized 
by a  best fit stellar atmosphere model (blue stars) of T = 9250 K, and an absolute K band magnitude of $–10.9$. 
The extinction and absolute magnitude are derived using the  Galactic center interstellar extinction law  
of \citet{nishiyama09}.  
It may be that this source is intrinsically hot, but we  can not rule out the  possibility that there 
is  a systematic error in the temperature derivation. This error could result from using  a subset of the available model atmosphere parameters 
or  the possible presence of warm dust that adds to the stellar emission at wavelengths as short as 3.4 $\mu$m.
The high temperature   and the luminosity  could  be matched by a young massive star. 

As the northern extension of the X-ray filament from position A crosses the cluster of stars, 
it does not follow the eastern radio filament.  
To demonstrate  this, Figure 4 compares the intensity in radio and X-rays 
along three background subtracted slices across the 
width of the filaments, as labeled by horizontal lines  on an  X-ray image. 
The eastern filament has an X-ray counterpart. However, the separation of the peak emission 
from the radio  filament with respect to its X-ray counterpart 
increases to the north is  shown in Figure 4b (slice 3).

\subsection{Radio and X-ray spectrum  of G0.173-0.42}

The spectral index  $\alpha$
of G0.173-0.42 is determined 
by measuring the flux density F$_\nu$  at   15  frequency channels within the broad 20 cm bandwidth, as shown in 
Figure 5a.
G0.173-0.42  shows a north-south  
gradient with the  steepest and flattest  spectral indices of  $\alpha\sim-1$ and 
 $\sim-0.2$  at  the northern and southern  ends of the parallel filaments, respectively.
The eastern and western filaments are closer together to  the south, as seen on Figures 1a and 2, where 
the flattest spectral index is noted.  Furthermore, the emission is strongest south of position A. 
It is possible that enhanced brightening and flatter 
spectral index are  due to a  new generation of energetic particles that are  accelerated 
at the interaction site, as discussed below.  
 
In  a  close-up view of  the region near position A, 
a change in the spectral index is noted exactly where the X-ray filament lies. 
To illustrate this, 
Figure 5b shows the spectral index  determined from  a slice  cutting across 
the width of the filament. 
The eastern and western filaments show a flattening  of $\alpha\sim0.3-0.35$ with respect to background emission with steeper 
spectrum. The intensity of 1631.6 MHz  emission  from a narrow channel is also plotted in  Figure 5b. 

Figure 6 shows a plot of 
radio and X-ray flux and their corresponding spectral indices.
The integrated flux densities of the entire X-ray filament and the corresponding radio counterpart at 1.28 GHz are shown. 
The 
extrapolations of the radio and X-ray spectra are shown in red and blue, respectively, with dashed lines and shaded 
regions indicate the best fit spectral indices and their 90\% confidence intervals. 
The 90\% confidence level on the 
radio spectral index is computed from  --$1.18\pm0.16$  with a flux density of 75.5 mJy.

\section{Discussion}

The  increase  in the  X-ray and radio brightness of eastern filament at the location where the oblique 
filament  crosses G0.173-0.42  suggests that 
 the oblique filament is physically interacting with 
the parallel filaments. 
This is also supported by the observation that 
the oblique filament  becomes fainter and wider from SE to NW  as it crosses the parallel filaments.  

We also note that the spectral index of radio emission is flatter  to the south along the radio/X-ray 
filaments than to the north of the filament. This suggests 
that a new  population of GeV electrons is injected  where the oblique 
filament 
crosses the parallel filaments.

\subsection{A model of X-ray and radio emission}

\subsubsection{Galactic wind}

Recent detection of  large scale X-ray and synchrotron emission
above and below the central molecular zone (CMZ) was interpreted as  arising from   cosmic-ray
driven outflow \citep{zadeh19}.
In this picture, the cosmic-ray momentum and energy are mediated by the
magnetic field and transferred to accelerating and heating the gas. The
cosmic rays and heated gas open a channel away from the Galactic plane and expand as a Galactic wind. 
The interaction of this wind with 
any compact  obstacles such as stellar wind bubbles creates the radio filaments 
which are analogous to  cometary tails behind mass-losing stars like Mira-type stars \citep{martin07}. 
In the case of 
G0.173-0.42, the Galactic center wind is expected to 
flow towards more negative latitudes,  away from the Galactic plane.
A schematic diagram of this picture is shown in Figure 7.

\subsubsection{Interaction of two  filaments: magnetic reconnection}

Because of the morphological evidence of the physical interaction of the oblique and north-south nonthermal radio 
filaments, it is plausible that reconnection of their magnetic field lines generates a new population of energetic 
particles.  This population is responsible not only for the sudden increase in radio emission but also X-ray 
emission from G0.173-0.42 at point A.

The 1.2\,GHz radio continuum emission along the eastern filament brightens to the South of point A, where it is crossed by the oblique 
filament.  This brightening is accompanied by the flattening of the spectrum, from $\sim \nu^{-1.4}$ to North of A, to $\nu^{-0.5}$ to the 
South.  The profile of the oblique filament becomes more diffuse and complex as it crosses the vertical filament
at A. Together, these changes point to a physical interaction of the filaments at position A suggestive 
of the acceleration of electrons at A. One 
possible form of this interaction  is magnetic reconnection. In this case,  some of the dissipated magnetic energy is transferred to a new population of relativistic 
electrons. To estimate the available magnetic power we adopt a nominal field strength $B=100$ \,$\mu$G (\cite{zadeh05,zadeh19}), 
reconnection speed $v=20$\,km/s, and 
cross-sectional area $A = (0.2\,\u pc )^2$.  Then the reconnection gives a power, 
$L_r \sim v_A\, A\, B^2/8\pi \approx 3\ee 32 \u erg\, \ut s -1 $, and 
the lifetime of the interaction is of order the filament width, 0.2\,pc, 
divided by the reconnection speed, 20 \kms, i.e. $\sim 10^4$ years.

The energy of the electrons dominating the emission at $1.2$\,GHz is $ \sim 0.9$\,GeV, and their synchrotron lifetime is $\sim 0.9$\,Myr.  The 
enhanced radio emission associated with this population extends $\approx 2$\,pc to the south. If this were determined by the synchrotron 
lifetime, then their propagation speed would only be $\sim 2$\,\kms, an order of magnitude below the $\sim 20$\,\kms cosmic-ray streaming 
speed that is commonly inferred elsewhere.  Instead, the spatial extent of the enhanced continuum emission is likely set by the time elapsed 
since the interaction between the filaments commenced.  If the total interaction time scale between the filament is $10^{4}$\,yr, as estimated 
above, then the age of the electrons must be less than this, suggesting a propagation speed in excess of 200\,\kms.  This is far higher than 
the speeds associated with cosmic-ray streaming.

This is consistent with a model in which non-thermal filaments at the Galactic center are a result 
of a large-scale cosmic-ray driven wind \citep{zadeh19}.  In this picture a net flow of plasma away from the 
galactic plane at hundreds of \kms is driven by the extreme cosmic-ray pressure in the central 150\,pc of the 
Galaxy \citep{oka19}.  The nonthermal filaments are magnetized streamers created by the wrapping of the wind's 
magnetic field around an obstacle to the flow, by analogy with the magnetized ion tails of comets embedded in the 
solar wind.  The relativistic electrons injected by reconnection are advected by this flow, and the time to be 
transported from the injection point A to the southern tip of the radio-brightened portion for the 
filament is a few thousand years, consistent with the expected interval since the interaction between the two 
filaments began.

\subsubsection{Enhanced X-ray emission: ICS}

We now turn to the origin of the X-ray emission.  This is unlikely to be synchrotron emission for three reasons.  
First, the distribution does not match that of the brighter, flat spectrum radio emission, but extends a couple of pc 
to the north and south of the injection point A.  Second, the lifetime of the $\sim 30$\,TeV electrons that would 
radiate in the keV range via synchrotron emission would be only 30 years, implying that their propagation speed to the 
north and south from the injection point would need to exceed $0.2\,c$. Third, the X-ray flux is too high to be 
consistent with the spectral index at radio wavelengths.  
To support this, we determined the integrated 1.28 GHz flux 
with a beam size of 8$''\times8''$ over the area covering the X-ray filament, $\sim 25.2''\times167.5''$, giving 
75.5$\pm13.2$ mJy.  The spectral index in the 20 cm band  is estimated to be --$1.18\pm0.16$ and is shown in Figure 6.

Instead an inverse-Compton scattering (ICS) scenario seems preferred, in which the X-ray emission is created by the upscattering of seed 
photons contributed by the bright stellar source located at the northern tip of the X-ray emission. The ICS flux $F_{\nu_x}$ at frequency 
$\nu_x$ is given by

\begin{equation}
 \nu_x F_{\nu_x}\approx 0.5\;\frac{\nu_1 U_{\nu_1}}{U_B}\,\nu_0 F_{\nu_0}
    \label{eqn:ICS}
\end{equation}

$\nu1\, U_{\nu_1} $ is the energy density of the seed photons at the peak of the stellar spectrum at frequency $\nu_1$, $U_B$ is the energy density 
magnetic field, and $\nu_0$ is the characteristic frequency of the synchrotron emission of electrons with Lorentz factor $\gamma \sim 
(\nu_X/\nu_1)^{1/2} $.

Adopting a stellar luminosity $L_\mathrm{*} \approx 1\ee 6 \lsol$, with an SED peaking 
at $\nu_1= 0.5\,\mu$m, typical of a young massive star, then upscattering to $h \nu_X 
\approx 5$\,keV is achieved by electrons in the low energy tail with Lorentz factors 
$\approx 50$.  The energy density at distance $d$ from the star is $U_\nu = L_\nu 
/(4 \pi c d\,^2)$.  
 Adopting a 
synchrotron flux of 0.1\,Jy at 1.2\,GHz and a $\nu^{-1.4}$ spectrum then $\nu F_\nu 
\propto \nu^{-0.4}$ and for our nominal field strength of 100\,$\mu$G 
(\cite{zadeh05,zadeh19}), 
this yields 
0.5--10 keV  X-ray flux over the extent of the X-ray filament of about 
$2\ee -13 $ 
erg\, cm$^{-2}$\, s$^{-1}$, 
consistent with the observed X-ray flux. Reducing the 
adopted magnetic field strength to 50 $\mu$G increases the X-ray flux to $8\ee -13 $
erg\, cm$^{-2}$\, s$^{-1}$.

In this scenario the X-ray emission traces $\sim$20 MeV electrons that are injected at (a) 
the stellar bow shock, and (b) the reconnection point where the oblique filament 
crosses the north-south filament at point A (see Fig. 6).  
In both cases the injected particles are 
carried southwards by the large scale nuclear wind from the inner 100 pc of the Galaxy.  
Inverse Compton scattering off the additional electrons injected at A offsets the 
$1/r^2$ decline of the seed photon density with distance from the star and maintains a 
relatively uniform surface brightness along the north-south filament.

\newcommand{\pyr}{yr$^{-1}$}

\subsection{Summary} 

We have presented a moderate  resolution study of the 
radio structure and spectral index of G0.173-0.42 which consists of two parallel filaments and   an oblique 
filament G0.167-0.405
crossing the parallel filaments. A small portion of one of the two parallel filaments has an X-ray counterpart. 
Noting  the variation in spectral index, sudden 
brightening and asymmetry of radio and X-ray emission, we argued for an interaction of the 
 oblique filament with one of the two parallel 
filaments and suggested injection of  a new population of relativistic particles due to magnetic field reconnection. 
The presence of a Galactic center wind produces lopsided profiles through its tendency 
to push particles away from the injection point.  We suggested  that the inverse Compton scattering mechanism 
is more likely to explain the X-ray emission by upscattering  of seed photons emitted from a  $10^6$ \lsol\, star located  at the northern tip of the X-ray filament. 
We  argued against  the synchrotron mechanism  for  the 
X-ray emission  due to the short $\sim$30 year lifetime of TeV relativistic particles.

\section{Data Availability}

All the data including  X-ray, radio and IR data  that we used here are available online and are not proprietary.
 We have reduced and calibrated these data and are available if  requested.

\section*{Acknowledgments}
We thank the referee for useful comments. This work is partially supported by the grant AST-0807400 from the
National  Science Foundation.
The MeerKAT telescope is operated by the South African Radio Astronomy Observatory, which is a facility of the National Research Foundation, an agency of the Department of Science and Innovation.
IH acknowledges support from the UK Science and Technology Facilities Council [ST/N000919/1], the Oxford Hintze Centre for Astrophysical Surveys which is funded through generous support from the Hintze Family Charitable Foundation, and a visiting Professorship from SARAO. We acknowledge use of the 
Inter-University Institute for Data Intensive Astronomy (IDIA) data intensive research cloud for data processing. IDIA is a South African university partnership involving the University of Cape Town, the University of Pretoria and the University of the Western Cape. MW thanks the Cherrybrook Research Institute for 
hospitality. The authors acknowledge the Centre for High Performance Computing (CHPC), South Africa, for providing computational resources to this research project.




\bibliographystyle{mnras}

\onecolumn

\begin{figure}
\centering
\includegraphics[scale=.95,angle=0]{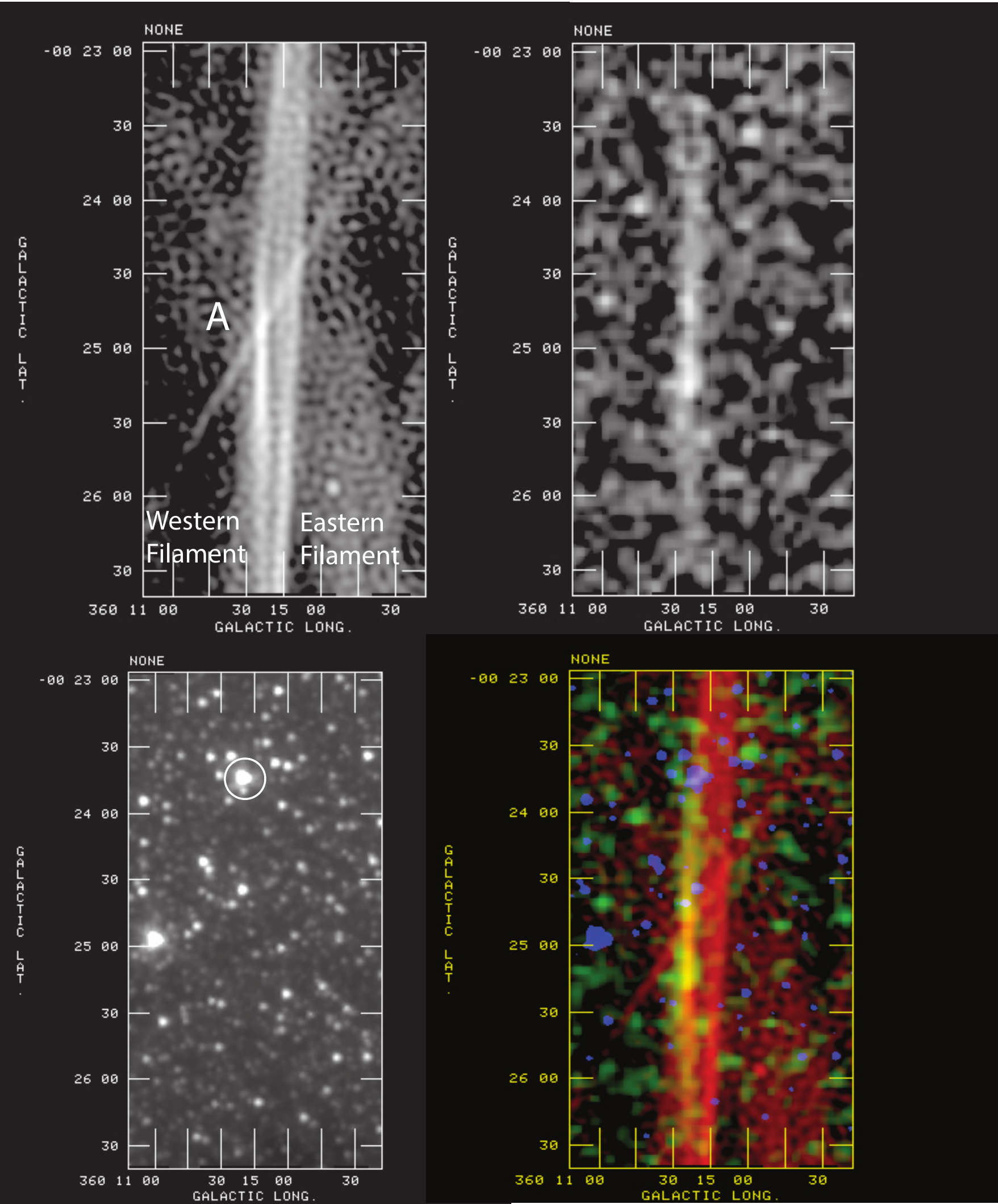}
\caption{
{\it a(Top Left)}
A grayscale MeerKAT image at 20 cm showing an oblique filament crossing two  parallel filaments 
at position A with a resolution of 
$\sim4''\times4''$. The eastern and western filaments are labeled. 
Note that diffuse radio emission surrounds the filaments.
{\it b(Top Right)}
A {\it Chandra} X-ray image integrated between  2-8 keV shows a single filament coincident with the 
eastern radio filament but  its northern extension from position A deviates to the NW in the direction away 
from the eastern radio filament. A circles shows the position of the bright star shown in (c). 
{\it c(Bottom Left)}
A 3.6 $\mu$m image of G0.173-0.42 taken with {\it Spitzer}/IRAC of {\it Spitzer}. 
A  bright saturated star is located where the X-ray filament 
splits into  two components. The bright saturated star is marked  with a circle. 
{\it d(Bottom Right)}
A composite RGB color image of the G0.173-0.42 filament at radio (R) and X-ray (G) 
superimposed on an IRAC image (B) at 3.6 $\mu$m. 
}
\end{figure}

\begin{figure}
\centering
\includegraphics[scale=0.8,angle=0]{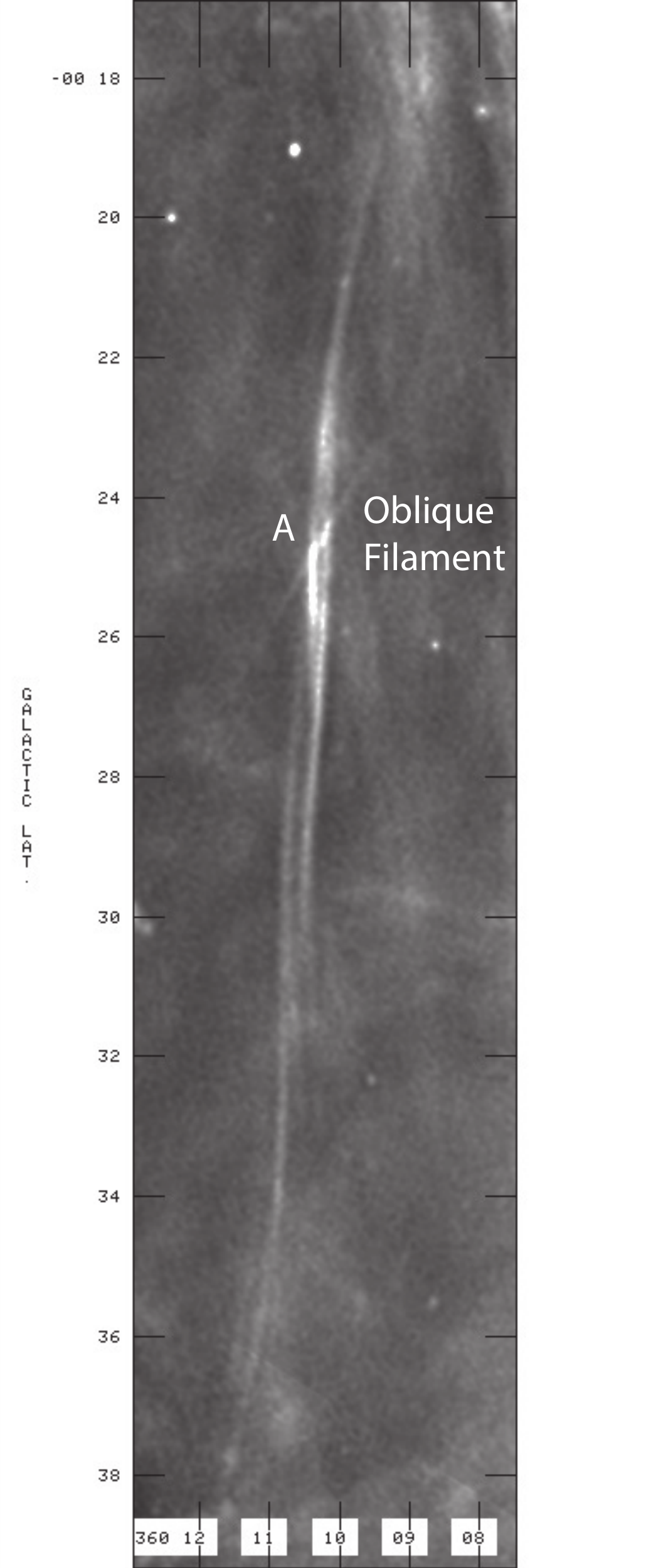}
\caption{A 1.28 GHz image of a network of parallel filaments, the brightest portion of which 
includes G0.173-0.42. 
The range of intensity ranges  between  
$-5\times10^{-5}$  and  3$\times10^{-4}$ Jy\, beam$^{-1}$ and the spatial resolution is $\sim6''$. 
}
\end{figure}

\begin{figure}
\centering
\includegraphics[scale=1.0,angle=0]{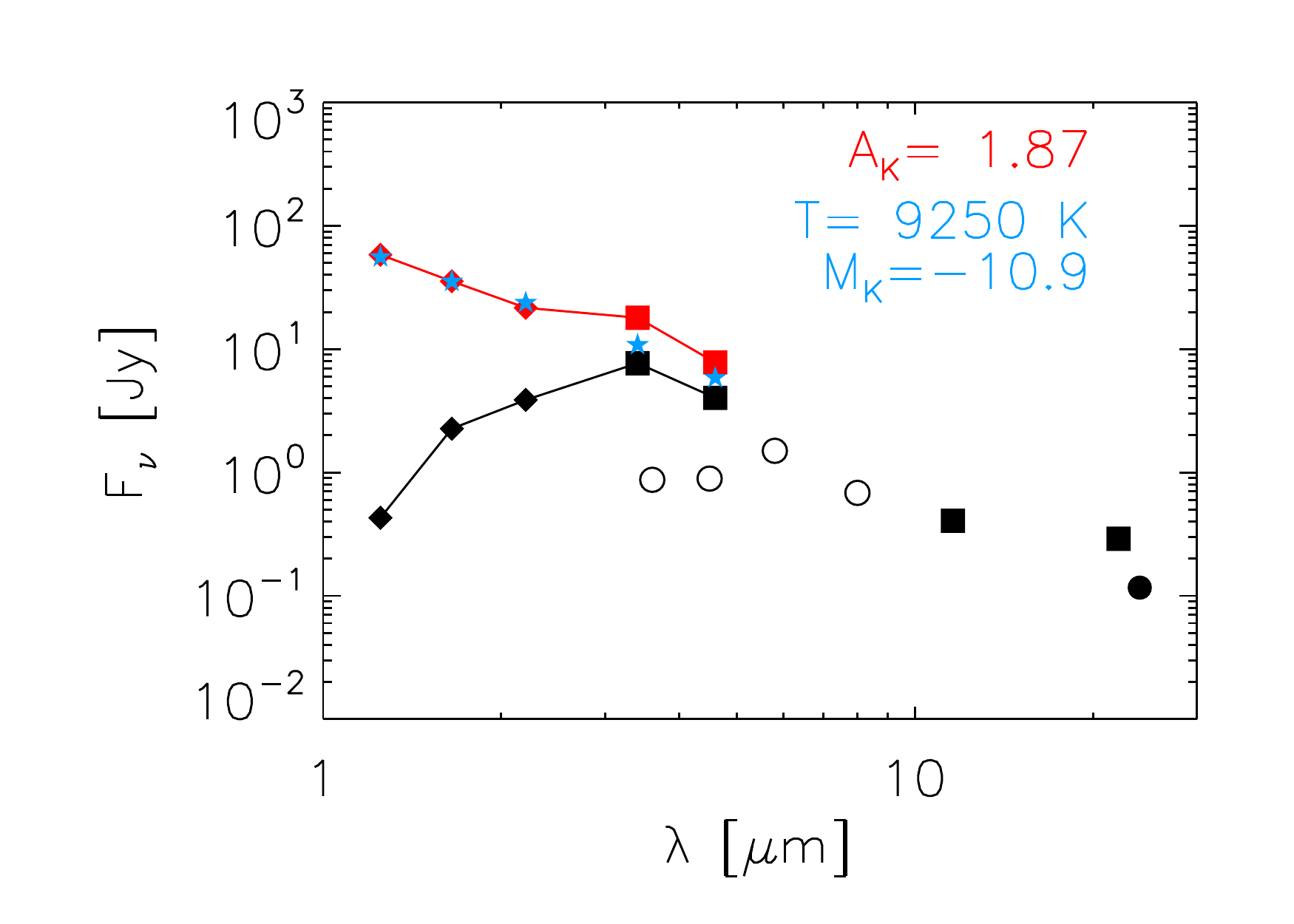}
\caption{
A spectral energy distribution (SED) of the brightest star in the cluster at the
northern tip of the X-ray filament. Observed flux densities are designated by
black symbols: 2MASS (diamonds), IRAC (likely affected by saturation, open
circles), MIPS (filled circle) and {\it WISE} (squares). Red symbols indicate
the extinction-corrected SED derived from a weighted fit between the 
2MASS + {\it WISE} data and the \citet{coelho14} stellar models. The fit indicates
an extinction of $A_K = 1.87$ mag, and a stellar atmosphere model (blue stars)
with $T = 9250$ K and an absolute K band magnitude of 
$M_{K} = -10.9$.
}
\end{figure}

\begin{figure}
\centering
\includegraphics[scale=0.7,angle=0]{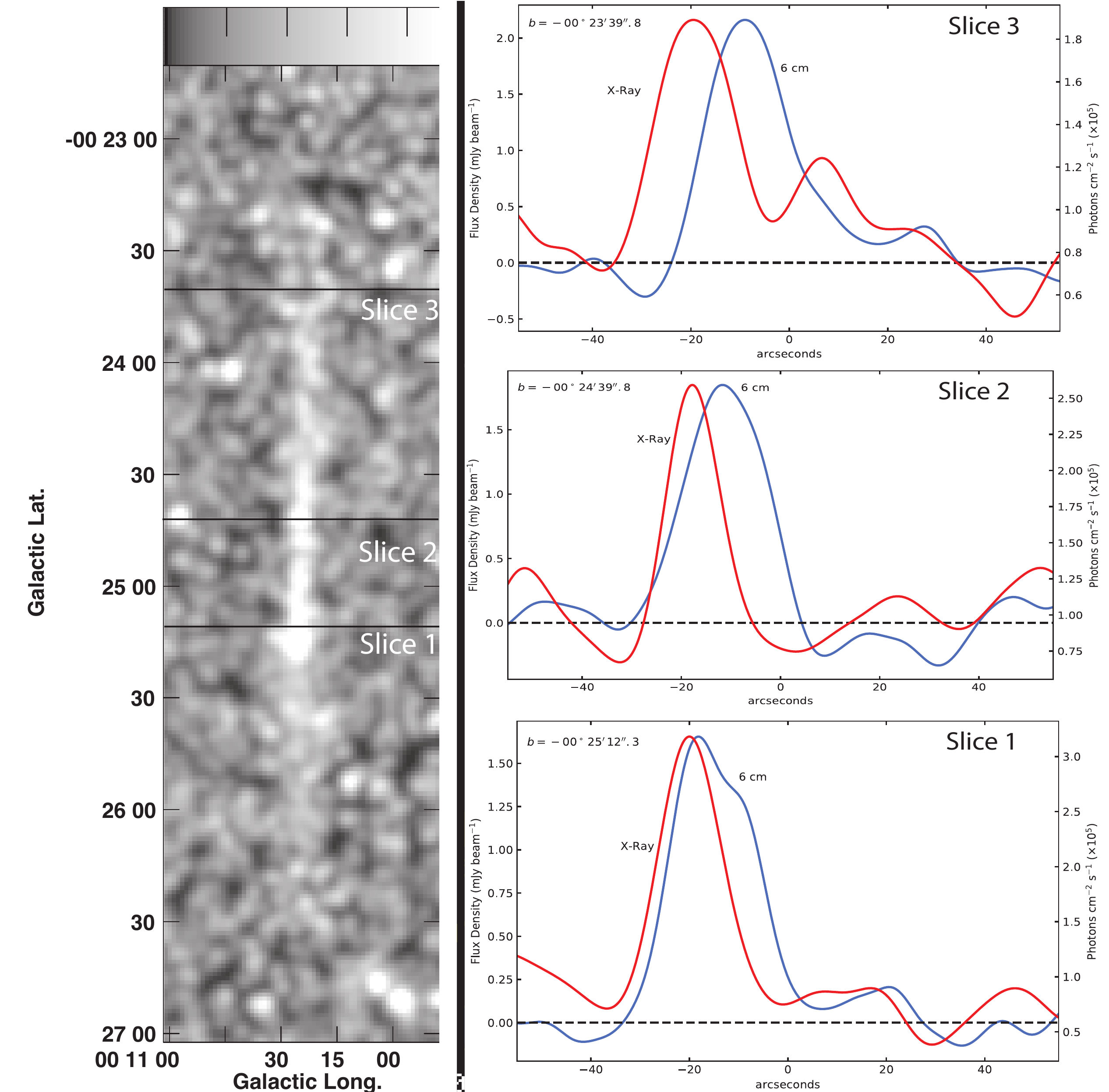}
\caption{
{\it a(Left)}
Lines on the X-ray image show the locations of profile measurements.  
Only the eastern filament of G0.173-0.42 appears in X-rays. 
The parallel western radio filament is $~10''$ away.
{\it b(Right  three panels)} The slice profiles show that the 
eastern radio filament shows an X-ray counterpart. The X-ray flux is integrated between 
1.5 and 5 keV. Both radio and X-ray data are convolved to a Gaussian beam $11''.95\times8''.69$ with position angle 
(PA) 72$^\circ$.55. 
Linear fits were used to subtract background emission from these profiles.
}
\end{figure}

\begin{figure}
\centering
\includegraphics[scale=0.7,angle=0]{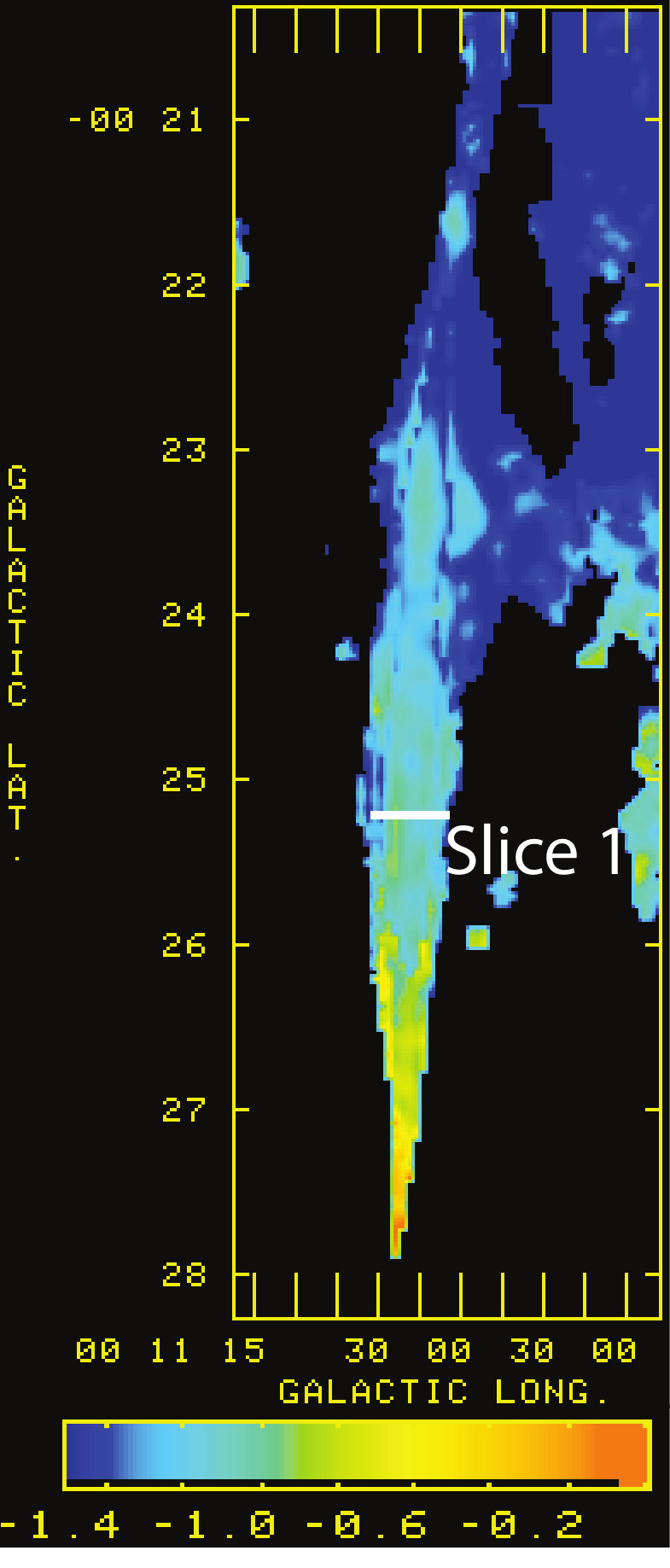}
\includegraphics[scale=0.6,angle=0]{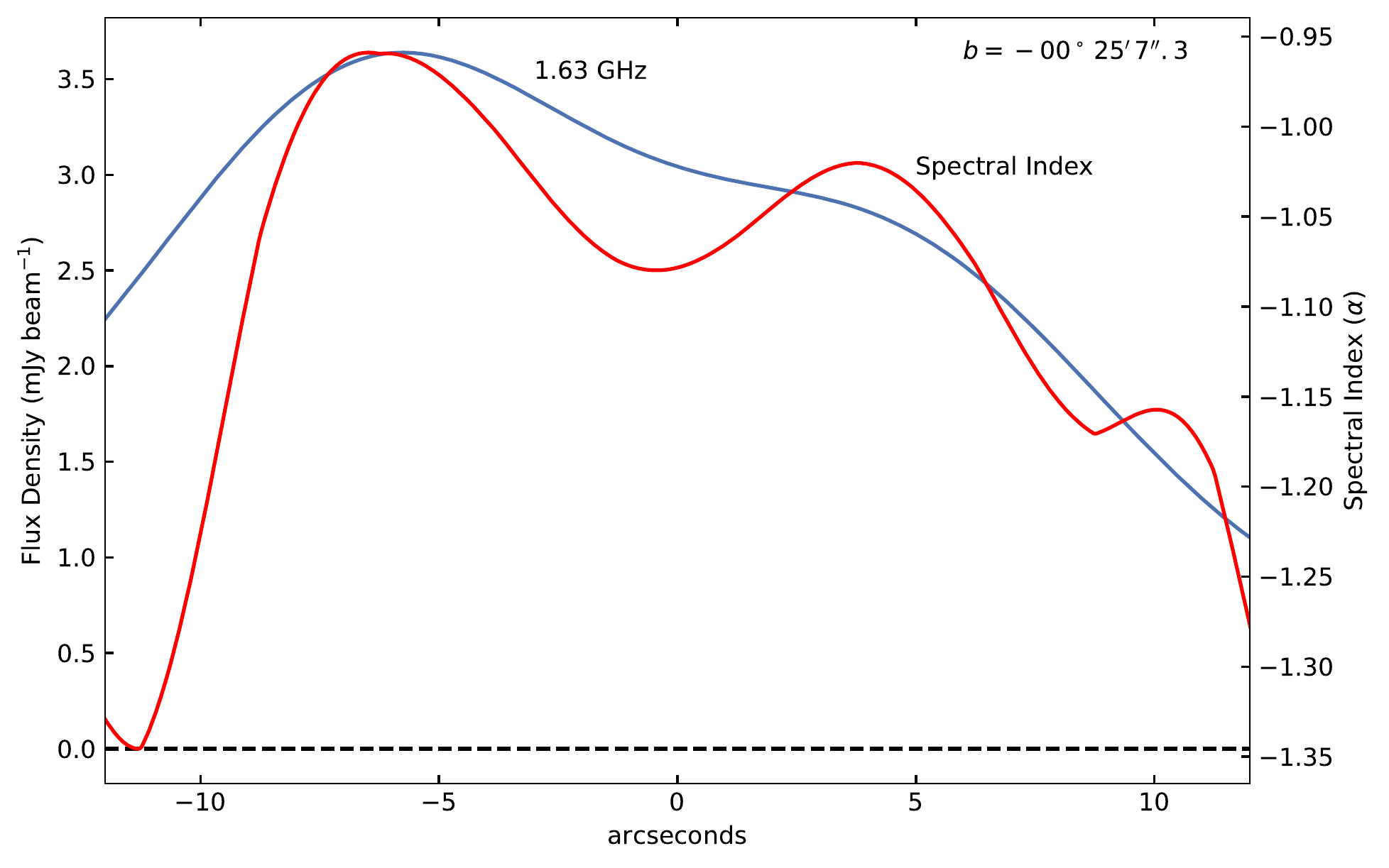}
\caption{
{\it a(Left)}
A spectral index image based on 15 MeerKAT sub-bands within the 20 cm band between 0.8 to 1.6 GHz  
convolved with an 8$''$ resolution. 
{\it b(Right)} 
The spectral index  (red-line) determined from  a slice  cutting across 
the width of the filament, as labeled by a horizontal line segment on Figure 5a. 
The 1631.6 MHz intensity profile (blue line) of the filaments is shown for reference.}
\end{figure}

\begin{figure}
\centering
\includegraphics[scale=0.5,angle=0]{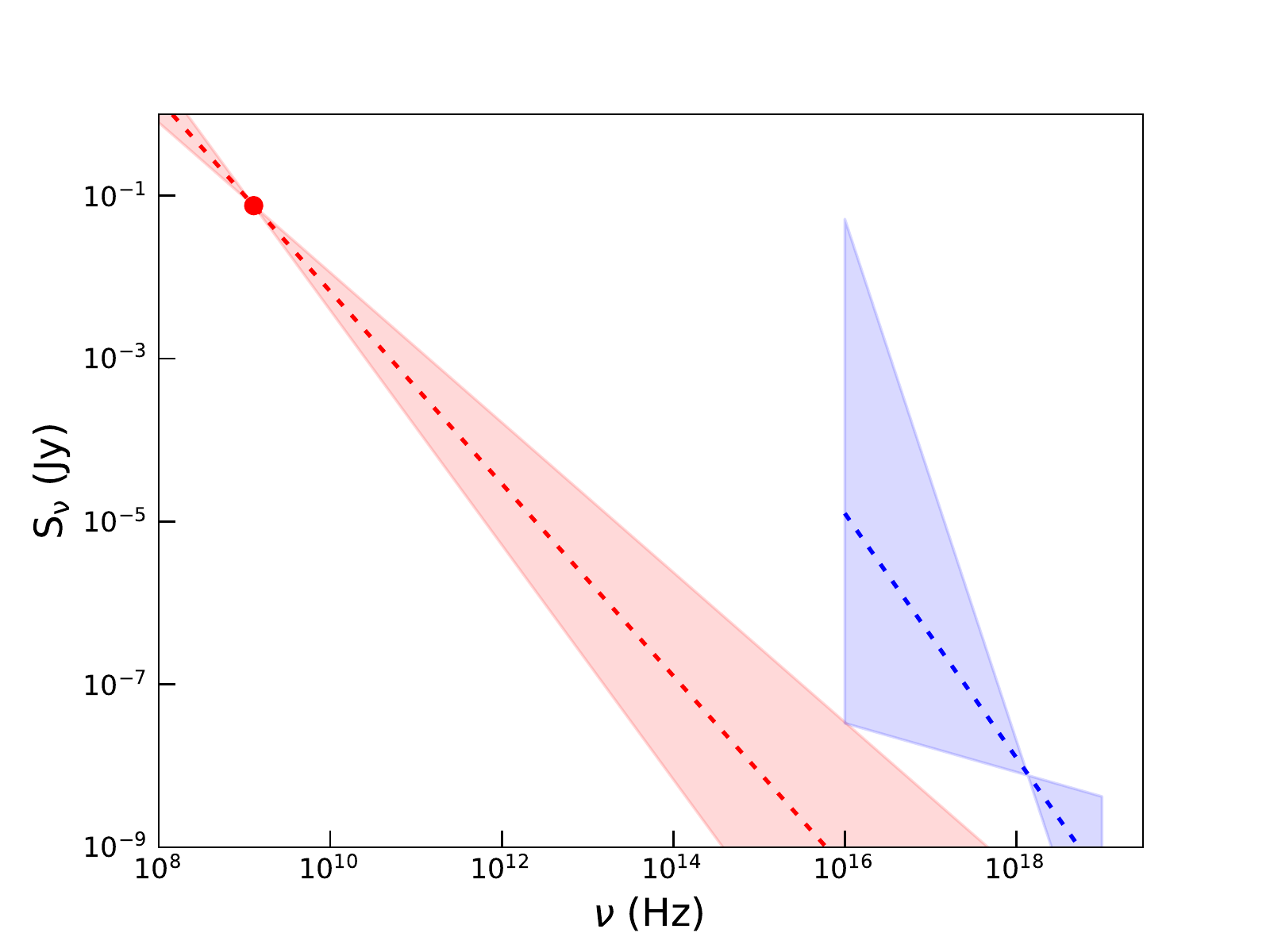}
\caption{
The integrated flux densities of the entire X-ray filament and the corresponding radio counterpart at 1.28 GHz. The 
extrapolations of the radio and X-ray spectra are shown in red and blue, respectively, with dashed lines and shaded 
regions indicate the best fit spectral indices and their 90\% confidence intervals. The 90\% confidence level on the 
radio spectral index is computed from  --$1.18\pm0.16$  with a flux density of 75.5 mJy.
took them from the text. The X-ray photon index 
 $\Gamma=2.5^{+1.7}_{-1.2}$  with  a 2-10 keV fluxes (unabsorbed)   
$2.0^{+2.5}_{-0.5}\times10^{-13}$ ergs s$^{-1}$ are used to determine the best fit spectral index to X-ray data. 
}
\end{figure}

\begin{figure}
\centering
\includegraphics[scale=0.6,angle=0]{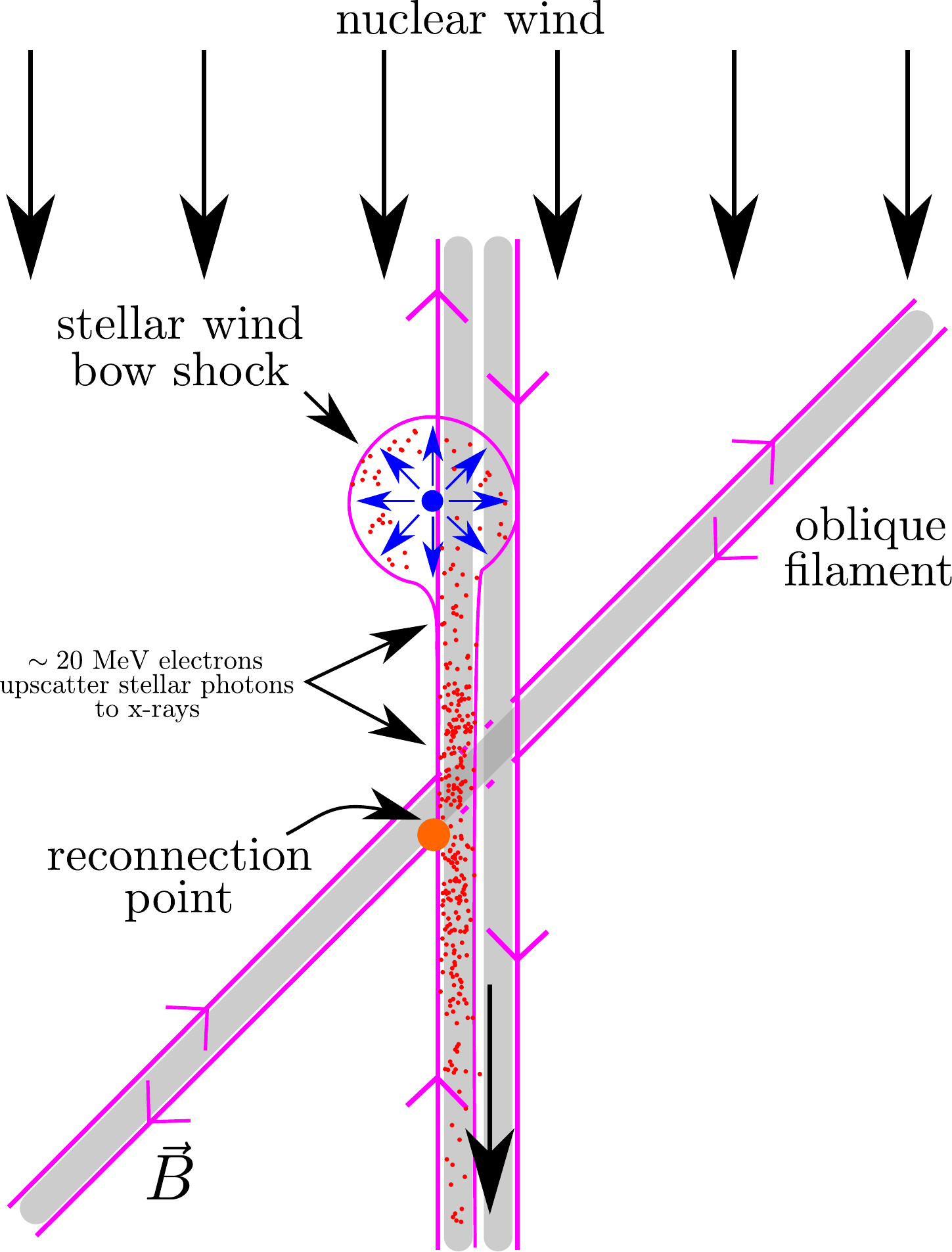}
\caption{
A schematic diagram of an oblique filament crossing two parallel filaments, producing 
a new population of relativistic particles. Cosmic ray 
particles from the nuclear wind and the reconnection point 
interact  in two locations with the seed photons of a 
luminous mass-losing star and produce X-ray emission due to ICS, as shown in red.  
}
\end{figure}

\end{document}